%Paper: hep-th/9309041
%From: POPE@PHYS.TAMU.EDU
%Date: Tue, 7 Sep 1993 16:37:00 CDT

%%%%%%%%%%%%%%%%%%%%%%%%%%%%%%%%%%%%%%%%%%%%%%%%%%%%%%%%%%%%%
%%%                                                       %%%
%%%        THE COMPLETE COHOMOLOGY OF THE W3 STRING       %%%
%%%                                                       %%%
%%%           H. Lu, C.N. Pope X.J. Wang and K.W. Xu      %%%
%%%                                                       %%%
%%%                    USE PLAIN TEX                      %%%
%%%                                                       %%%
%%%%%%%%%%%%%%%%%%%%%%%%%%%%%%%%%%%%%%%%%%%%%%%%%%%%%%%%%%%%%

\def\singlespace{\normalbaselines}
\def\oneandahalfspace{\baselineskip=1.15\normalbaselineskip plus 1pt
\lineskip=2pt\lineskiplimit=1pt}

\def\np{\vfill\eject}
\def\nl{\hfil\break}

\def\nofirstpagenoten{\nopagenumbers\footline={\ifnum\pageno>1\tenrm
\hss\folio\hss\fi}}
\def\nofirstpagenotwelve{\nopagenumbers\footline={\ifnum\pageno>1\twelverm
\hss\folio\hss\fi}}
\def\leaderfill{\leaders\hbox to 1em{\hss.\hss}\hfill}
\def\ft#1#2{{\textstyle{{#1}\over{#2}}}}
\def\frac#1/#2{\leavevmode\kern.1em
\raise.5ex\hbox{\the\scriptfont0 #1}\kern-.1em/\kern-.15em
\lower.25ex\hbox{\the\scriptfont0 #2}}
\def\sfrac#1/#2{\leavevmode\kern.1em
\raise.5ex\hbox{\the\scriptscriptfont0 #1}\kern-.1em/\kern-.15em
\lower.25ex\hbox{\the\scriptscriptfont0 #2}}

  %20 point
                   %17 point
  %14 point
 %17 point
 %14 point
 %14 point
 %14 point

\parindent=20pt
\def\narrow{\advance\leftskip by 40pt \advance\rightskip by 40pt}

\def\AB{\bigskip
        \centerline{\bf ABSTRACT}\medskip\narrow}
\def\nonarrower{\advance\leftskip by -40pt\advance\rightskip by -40pt}
\def\AE{\bigskip\nonarrower}

\def\boxit#1{\vbox{\hrule\hbox{\vrule\kern3pt
        \vbox{\kern3pt#1\kern3pt}\kern3pt\vrule}\hrule}}

\def\gtorder{\mathrel{\raise.3ex\hbox{$>$}\mkern-14mu
             \lower0.6ex\hbox{$\sim$}}}
\def\ltorder{\mathrel{\raise.3ex\hbox{$<$}|mkern-14mu
             \lower0.6ex\hbox{\sim$}}}
\def\dalemb#1#2{{\vbox{\hrule height .#2pt
        \hbox{\vrule width.#2pt height#1pt \kern#1pt
                \vrule width.#2pt}
        \hrule height.#2pt}}}

\font\fourteentt=cmtt10 scaled \magstep2
\font\fourteenbf=cmbx12 scaled \magstep1
\font\fourteenrm=cmr12 scaled \magstep1
\font\fourteeni=cmmi12 scaled \magstep1
\font\fourteenss=cmss12 scaled \magstep1
\font\fourteensy=cmsy10 scaled \magstep2
\font\fourteensl=cmsl12 scaled \magstep1
\font\fourteenex=cmex10 scaled \magstep2
\font\fourteenit=cmti12 scaled \magstep1
\font\twelvett=cmtt10 scaled \magstep1 \font\twelvebf=cmbx12
\font\twelverm=cmr12 \font\twelvei=cmmi12
\font\twelvess=cmss12 \font\twelvesy=cmsy10 scaled \magstep1
\font\twelvesl=cmsl12 \font\twelveex=cmex10 scaled \magstep1
\font\twelveit=cmti12
\font\tenss=cmss10
 
 \font\ninebf=cmbx7 scaled \magstep1
\font\ninerm=cmr7 scaled \magstep1 \font\ninei=cmmi7 scaled \magstep1
\font\ninesy=cmsy7 scaled \magstep1 
\font\eightrm=cmr7 scaled 1140 
 
\font\sevenbf=cmbx7 \font\sevenrm=cmr7 \font\seveni=cmmi7
\font\sevensy=cmsy7 

\catcode`@=11
\newskip\ttglue
\newfam\ssfam

\def\fourteenpoint{\def\rm{\fam0\fourteenrm}
\textfont0=\fourteenrm \scriptfont0=\tenrm \scriptscriptfont0=\sevenrm
\textfont1=\fourteeni \scriptfont1=\teni \scriptscriptfont1=\seveni
\textfont2=\fourteensy \scriptfont2=\tensy \scriptscriptfont2=\sevensy
\textfont3=\fourteenex \scriptfont3=\fourteenex \scriptscriptfont3=\fourteenex
\def\it{\fam\itfam\fourteenit} \textfont\itfam=\fourteenit
\def\sl{\fam\slfam\fourteensl} \textfont\slfam=\fourteensl
\def\bf{\fam\bffam\fourteenbf} \textfont\bffam=\fourteenbf
\scriptfont\bffam=\tenbf \scriptscriptfont\bffam=\sevenbf
\def\tt{\fam\ttfam\fourteentt} \textfont\ttfam=\fourteentt
\def\ss{\fam\ssfam\fourteenss} \textfont\ssfam=\fourteenss
\tt \ttglue=.5em plus .25em minus .15em
\normalbaselineskip=16pt
\abovedisplayskip=16pt plus 4pt minus 12pt
\belowdisplayskip=16pt plus 4pt minus 12pt
\abovedisplayshortskip=0pt plus 4pt
\belowdisplayshortskip=9pt plus 4pt minus 6pt
\parskip=5pt plus 1.5pt
\setbox\strutbox=\hbox{\vrule height12pt depth5pt width0pt}
\let\sc=\tenrm
\let\big=\fourteenbig \normalbaselines\rm}
\def\fourteenbig#1{{\hbox{$\left#1\vbox to12pt{}\right.\n@space$}}}

\def\twelvepoint{\def\rm{\fam0\twelverm}
\textfont0=\twelverm \scriptfont0=\ninerm \scriptscriptfont0=\sevenrm
\textfont1=\twelvei \scriptfont1=\ninei \scriptscriptfont1=\seveni
\textfont2=\twelvesy \scriptfont2=\ninesy \scriptscriptfont2=\sevensy
\textfont3=\twelveex \scriptfont3=\twelveex \scriptscriptfont3=\twelveex
\def\it{\fam\itfam\twelveit} \textfont\itfam=\twelveit
\def\sl{\fam\slfam\twelvesl} \textfont\slfam=\twelvesl
\def\bf{\fam\bffam\twelvebf} \textfont\bffam=\twelvebf
\scriptfont\bffam=\ninebf \scriptscriptfont\bffam=\sevenbf
\def\tt{\fam\ttfam\twelvett} \textfont\ttfam=\twelvett
\def\ss{\fam\ssfam\twelvess} \textfont\ssfam=\twelvess
\tt \ttglue=.5em plus .25em minus .15em
\normalbaselineskip=14pt
\abovedisplayskip=14pt plus 3pt minus 10pt
\belowdisplayskip=14pt plus 3pt minus 10pt
\abovedisplayshortskip=0pt plus 3pt
\belowdisplayshortskip=8pt plus 3pt minus 5pt
\parskip=3pt plus 1.5pt
\setbox\strutbox=\hbox{\vrule height10pt depth4pt width0pt}
\let\sc=\ninerm
\let\big=\twelvebig \normalbaselines\rm}
\def\twelvebig#1{{\hbox{$\left#1\vbox to10pt{}\right.\n@space$}}}

\def\tenpoint{\def\rm{\fam0\tenrm}
\textfont0=\tenrm \scriptfont0=\sevenrm \scriptscriptfont0=\fiverm
\textfont1=\teni \scriptfont1=\seveni \scriptscriptfont1=\fivei
\textfont2=\tensy \scriptfont2=\sevensy \scriptscriptfont2=\fivesy
\textfont3=\tenex \scriptfont3=\tenex \scriptscriptfont3=\tenex
\def\it{\fam\itfam\tenit} \textfont\itfam=\tenit
\def\sl{\fam\slfam\tensl} \textfont\slfam=\tensl
\def\bf{\fam\bffam\tenbf} \textfont\bffam=\tenbf
\scriptfont\bffam=\sevenbf \scriptscriptfont\bffam=\fivebf
\def\tt{\fam\ttfam\tentt} \textfont\ttfam=\tentt
\def\ss{\fam\ssfam\tenss} \textfont\ssfam=\tenss
\tt \ttglue=.5em plus .25em minus .15em
\normalbaselineskip=12pt
\abovedisplayskip=12pt plus 3pt minus 9pt
\belowdisplayskip=12pt plus 3pt minus 9pt
\abovedisplayshortskip=0pt plus 3pt
\belowdisplayshortskip=7pt plus 3pt minus 4pt
\parskip=0.0pt plus 1.0pt
\setbox\strutbox=\hbox{\vrule height8.5pt depth3.5pt width0pt}
\let\sc=\eightrm
\let\big=\tenbig \normalbaselines\rm}
\def\tenbig#1{{\hbox{$\left#1\vbox to8.5pt{}\right.\n@space$}}}
\let\rawfootnote=\footnote \def\footnote#1#2{{\rm\parskip=0pt\rawfootnote{#1}
{#2\hfill\vrule height 0pt depth 6pt width 0pt}}}

\def\tenfoot{\tenpoint\hskip-\parindent\hskip-.1cm}

\overfullrule=0pt
\twelvepoint
\def\sbullet{\raise.2em\hbox{$\scriptscriptstyle\bullet$}}
\nofirstpagenotwelve
%\magnification=\magstep1%
\hsize=16.5 truecm
%\vsize=23.0 truecm
\baselineskip 15pt
%\parskip 0pt
%\font\ftf=cmr8

\def\ft#1#2{{\textstyle{{#1}\over{#2}}}}
\def\sss{\scriptscriptstyle}
\def\ket#1{\big| #1\big\rangle}

\def\del{\partial}

\oneandahalfspace
\rightline{CTP TAMU--50/93}
\rightline{hep-th/9309041}
\rightline{September 1993}

\vskip 2truecm
\centerline{\bf The Complete Cohomology of the $W_3$ String}
\vskip 1.5truecm
\centerline{H. Lu, C.N. Pope,\footnote{$^*$}{\tenfoot Supported in part
by the U.S. Department of Energy, under
grant DE-FG05-91ER40633.}
X.J. Wang and K.W. Xu$^*$}
\vskip 1.5truecm
\centerline{\it Center
for Theoretical Physics,
Texas A\&M University,}
\centerline{\it College Station, TX 77843--4242, USA.}

\vskip 1.5truecm
\AB\singlespace
We present a simple procedure for constructing the complete cohomology of
the BRST operator of the two-scalar and multi-scalar $W_3$ strings.  The
method consists of obtaining two level--15 physical operators in the
two-scalar $W_3$ string that are invertible, and that can normal order with all
other physical operators.  They can be used to map all physical operators into
non-trivial physical operators whose momenta lie in a fundamental unit cell.
By carrying out an exhaustive analysis of physical operators in this cell,
the entire cohomology problem is solved.
\AE
\oneandahalfspace

\np
\noindent
{\bf 1. Introduction}
\bigskip
Determining the physical spectrum of the $W_3$ string is equivalent to finding
the cohomology of its BRST operator. We shall be concerned with the BRST
operator of the critical $W_3$ string in this paper. The form of this
operator in terms of the abstract $W_3$ algebra was first given in [1].  In
the $W_3$ string, one realises the $W_3$ algebra in terms of a set of $n\ge
2$ free scalar fields [2,3], and the BRST operator takes the form [4]:
$$
\eqalignno{
Q_B&= Q_0 + Q_1,&(1.1)\cr
Q_0&=\oint dz\, c \Big(T^{\rm eff} +T_{\varphi} + T_{\gamma,\beta} + \ft12
T_{c,b} \Big), &(1.2)\cr
Q_1&=\oint dz\, \gamma\Big( (\del\varphi)^3 + 3\alpha\, \del^2\varphi\, \del
\varphi + \ft{19}8 \del^3\varphi +\ft92 \del\varphi\, \beta\, \del\gamma
+\ft32 \alpha\, \del\beta\, \del\gamma\Big),&(1.3)\cr}
$$
where the energy-momentum tensors are given by
$$
\eqalignno{
T_\varphi&\equiv -\ft12 (\del\varphi)^2 -\alpha\, \del^2\varphi, &(1.4)\cr
T_{\gamma,\beta}&\equiv -3\, \beta\,\del\gamma -2\, \del\beta\, \gamma,
&(1.5)\cr
T_{c,b}&\equiv -2\, b\, \del c - \del b\, c, &(1.6)\cr
T^{\rm eff} &\equiv -\ft12 \del X^\mu\, \del X^\nu\, \eta_{\mu\nu} -
i a_\mu\, \del^2 X^\mu. &(1.7)\cr}
$$
The background charge $\alpha$ for the scalar $\varphi$ is given by
$\alpha^2=\ft{49}8$, and the background-charge vector $a_\mu$ for $X^\mu$ is
chosen so that $d-12a_\mu \, a^\mu =\ft{51}2$.  The BRST operator is graded,
with $Q_0^2=Q_1^2 =\{Q_0,Q_1\}=0$. Note that the BRST operator we are using
here is the simplified one obtained in [4] by performing a non-linear field
redefinition involving $\varphi$ and the ghost fields.

In addition to the general $n>2$ multi-scalar string we shall be interested
especially in the two-scalar string. For this case we take the
energy-momentum tensor $T^{\rm eff}$ to be given by
$$
T^{\rm eff}=-\ft12 \big( \del X)^2- a\,\del^2 X\ ,\eqno(1.8)
$$
where $a^2=\ft{49}{24}$.

Many physical states were found for both the multi-scalar and two-scalar
$W_3$ strings in [4--8], by directly solving the physical-state conditions
$$
Q_B \ket{\chi} =0,\qquad  \ket{\chi}\ne Q_B \ket{\psi} \eqno(1.9)
$$
at particular level numbers and ghost numbers. The most extensive list of
such states can be found in [4].  From the examples that have
been constructed some suggestive patterns and structures have emerged,
especially in the case of the multi-scalar $W_3$ string, but such an
approach is necessarily not exhaustive.  A different approach was taken in
[9], where the cohomology of the non-critical $W_3$ BRST operator was studied
by  standard homological techniques. Results for the critical two-scalar $W_3$
string were obtained as a special case [9]. These results give the complete
cohomology for states whose $\varphi$ and $X$ momenta $(p_1,\,p_2)$ lie within
the ``Seiberg bounds;'' {\it i.e.}\ within a certain  wedge
($SU(3)$ Weyl chamber) in the
$(p_1,\,p_2)$ plane.\footnote{$^*$}{\tenfoot It should be emphasised that
neither in this case nor indeed for the ordinary Virasoro string is the
``Seiberg bound'' in any sense a restriction on the physical acceptability
of states in the free theory.  Rather, it is a bound on the values of
momentum that can occur in the exponential operators that approximate the
true wavefunctions of the interacting Toda or Liouville theory in the region
where the potential dies away.  Since there exists a many-to-one map (6--1
and 2--1 respectively) from the free theory into the interacting theory, a
condition analogous to the Seiberg bound arises as a statement that it
suffices to consider free-field operators with momenta within one Weyl
chamber in order to map into the entire Hilbert space of operators in the
interacting theory [10].  However, if one considers the free theory in its own
right then all physical operators are distinct, and there is no physical
equivalence between operators with momenta related by Weyl reflections.}

     In this paper we shall present a very simple and direct method that gives
the complete cohomology of the two-scalar critical $W_3$ BRST operator for
the entire $(p_1,\,p_2)$ plane. The method involves constructing certain
invertible physical operators which have the property that they normal order
with any physical operator to give another BRST non-trivial physical
operator. Since these invertible operators carry momentum, they, together
with their inverses, can be used to map any physical operator into one whose
momentum lies within a fundamental rectangular unit cell in the $(p_1,\,p_2)$
plane.  By carrying out an exhaustive analysis of all the physical states
whose momenta lie within this fundamental cell, it follows that the complete
cohomology problem for the BRST operator is then solved.

   We shall begin the discussion with the two-scalar $W_3$ string in
the next section. In section 3 we shall show how these results can be used
to determine the complete cohomology of the multi-scalar $W_3$
string. The same method can be applied to the study of the cohomology of the
one-scalar Virasoro string, which we shall do in section 4. We end with
discussion and conclusions in Section 5.

\bigskip
\noindent{\bf 2. The Cohomology of the Two-scalar $W_3$ String}
\medskip
\noindent{\it 2.1 Derivation of the cohomology}
\bigskip

       It was shown in [11] that the higher level physical states for the
two-scalar Virasoro string could be built by taking powers of two ground
ring generators $x$ and $y$. These operators have ghost number zero, which
is understandable when one considers that the physical operators have a
restricted range of ghost numbers $0\le G\le 3$, regardless of their level
number.  It is natural to expect that the physical states of the $W_3$
string should similarly be constructable in terms of some analogues of the
ground-ring generators.  However here the range of ghost numbers of the
physical operators is of the form $-r\le G\le r+8$, where $r$ increases
indefinitely with increasing level number.  Thus the natural analogues of
the ground-ring generators should have an appropriate negative ghost number.
They should  also have the property that they have well-defined normal-ordered
products  with all physical operators in the theory.

     The candidates that fulfil these criteria are two
physical operators with ghost number $G=-2$, at level $\ell=15$.  As we shall
show, these operators, which we call $x$ and $y$, have the remarkable
property that they have inverses $x^{-1}$ and $y^{-1}$, in the sense that
there are two physical operators $x^{-1}$ and $y^{-1}$ whose normal-ordered
products with $x$ and $y$ respectively yield the identity. This contrasts
with the situation for the two-scalar Virasoro string, where the ground ring
operators have no inverses.   In this section we shall begin by constructing
the $x$ and $y$ operators, and their inverses. Then we shall prove that
normal ordering $x$, $y$, $x^{-1}$ or $y^{-1}$ with any physical operator
will yield another BRST non-trivial physical operator. From this, we shall
be able to show that {\it all} physical operators in the two-scalar $W_3$
string can be written in the form $x^m\, y^n\, t_i$ and $x^m\, y^n\, u_i$,
where  $m$ and $n$ are any integers; $1\le i\le 6$; and $t_i$ and $u_i$ denote
sets of six ``basic'' physical operators at ghost numbers $3$ and $2$
respectively.

     All the known physical states of the two-scalar $W_3$ string have the
property that their momenta $(p_1,\, p_2)$ have the form
$$
(p_1,\, p_2)=(\ft17 k_1\, \alpha,\ \ft17 k_2\, a)\ ,\eqno(2.1)
$$
where $k_1$ and $k_2$ are integers, and $\alpha$ and $a$ are the background
charges for the $\varphi$ and $X$ directions, given by $\alpha^2=\ft{49}8$
and $a^2=\ft{49}{24}$.  It can be argued in general that the momenta of all
physical states must be of this form [12].\footnote{$^*$}{\tenfoot This
quantisation of the momenta as integer multiples of $\ft17\alpha$ and $\ft17
a$ is also seen in the results in [9].}  Thus it
is convenient to characterise the momentum of a physical state  by the pair of
integers $(k_1,\, k_2)$.  From (1.1)--(1.7) it is  straightforward to show
that $k_1$ and $k_2$ are related to the level number  $\ell$ by the mass-shell
condition [4] $$
3(k_1+7)^2 +(k_2+7)^2=4(12\ell+1)\ .\eqno(2.2)
$$
In particular, this places severe restrictions on the momenta, and level
numbers, that physical states can have.

     The $x$ and $y$ operators that we require turn out to arise as physical
states at level $\ell=15$, with momenta $(k_1,\, k_2)=(8,0)$ and $(4,12)$
respectively.  They have ghost number $G=-2$.  Finding their explicit form is
straightforward, although somewhat tedious; one simply writes down the most
general possible structure at the given level number and ghost number, and
then solves the physical-state conditions (1.9).  We made extensive use of
the Mathematica package OPEdefs [13] in order to do this.  In fact a
convenient simplification can be achieved by introducing the associated
screening currents $S_x$ and $S_y$, defined by
$$
S_x(z)\equiv \oint dw\, b(w)\, x(z)\eqno(2.3)
$$
(and similarly for $S_y$), in terms of which the physical-state condition
reduces to the equation $\{Q_B,S_x\}=\del x$ (and similarly for $y$).  The
result for the screening current associated with the $x$ operator is very
simple:
$$
S_x=\del^2\beta\, \del\beta\, \beta\, e^{\ft87 \alpha\varphi}\ .\eqno(2.4)
$$
The $x$ operator itself has 16 terms, and takes the form $x=\big( c\,
\del^2\beta\, \del\beta\, \beta + f(\varphi,\beta,\gamma)\big)\,
e^{\ft87\alpha\varphi}$; we need not give it explicitly  here.  The
screening current associated with the $y$ operator is much more  complicated,
and we shall not give it explicitly.  Unlike $S_x$, the screening current
$S_y$ involves the $b$ antighost, and the structure of $y$ is correspondingly
much more complicated than that of $x$ given above.  $S_y$ has the general
form
$$
S_y=\Big(\del^2\beta\, \del\beta\, \beta +\cdots \Big)  e^{\ft47\alpha\varphi
+\ft{12}7 a X}\ ,\eqno(2.5)
$$
where we have suppressed about fifty further terms.  The $y$ operator itself
has about 1000 terms.

     The physical operators that we shall call $x^{-1}$ and $y^{-1}$  are
much simpler. They are level 1 operators with ghost number $+2$, and
momenta $(k_1,\,k_2)=(-8,\,0)$ and $(-4,\,-12)$ respectively:
$$
\eqalignno{
x^{-1}&=\big(c\,\gamma+\ft{15}{14}\alpha\,\del\gamma\,\gamma\big)
e^{-\ft87\alpha\varphi}\ ,&(2.6)\cr
y^{-1}&=c\,\gamma\,e^{-\ft47\alpha\varphi-\ft{12}7 a X}\ .&(2.7)\cr}
$$
We have explicitly checked that the normal-ordered products $(x^{-1}\,x)$
and  $(y^{-1}\,y)$ give non-vanishing constant multiples of the identity
operator,  thus justifying the names $x^{-1}$ and $y^{-1}$. (The precise
values of the  non-vanishing constants are inessential to the subsequent
argument, and so  we shall assume renormalisations such that
$(x^{-1}\,x)=1$ and  $(y^{-1}\,y)=1$.)

       We now prove the following
\medskip
\noindent{\bf Lemma 2.1}\hskip10pt {\it If $A$ and $B$ are any two physical
operators that have a well-defined normal-ordered product $(A\,B)\equiv\oint
 {dz\over(z-w)} A(z)\,B(w)$, the commutator $[A,\,B]\equiv
(A\,B)-(-1)^{ab}(B\,A)$ is BRST trivial.}

To prove this, we note from [14] that
$$
[A,\,B]=(-1)^{ab} \sum_{r=1}^\infty\ {(-1)^r\over r!} \del^r\,[B\,A]_r ,
\eqno(2.8)
$$
where $[B\,A]_r$ denotes the coefficient of the $r$-th pole in the OPE of
$B$ with $A$.  (The sum in (2.8) terminates, since the degree of the highest
pole in the OPE of $B$ with $A$ is bounded.)  Thus $[A,\,B]=\del\eta$ for
some $\eta$. Since $A$ and $B$ are physical, it follows that $(A\,B)$ and
consequently $[A,\,B]$ are annihilated by the BRST operator $Q_B$.
Therefore $Q_B\,\del\eta=0$, implying $\del Q_B\,\eta=0$ and hence
$Q_B\,\eta=0$. Since $\del\eta=L_{-1}\eta=\{Q_B,\,b_{-1}\}\eta$, we
see that
$$
[A,\,B]=\del\eta= Q_B\,\xi\ ,\eqno(2.9)
$$
where $\xi=b_{-1}\eta$.

\bigskip

     The main results of this paper then depend crucially on the following
\medskip
\noindent{\bf Theorem 2.2}\hskip10pt {\it If $V$ is any BRST non-trivial
physical operator, then the normal-ordered products $(x\,V),\ (y\,V),\
(x^{-1}\,V)$ and $(y^{-1}\,V)$ are all BRST non-trivial physical operators.}

       We shall present the proof for $(x\,V)$; the proofs for the other cases
are identical.   That $(x\, V)$ is annihilated by $Q_B$ is obvious; what has to
be proved is that it is BRST non-trivial.  We begin by noting that for any
operators $A$, $B$ and $C$  [14],
$$
(A\,(B\,C))=(-1)^{ab} (B\,(A\,C)) + ([A,\,B]\,C)\ .\eqno(2.10)
$$
Thus we have
$$
(x^{-1}\, (x\, V))\approx ((x^{-1}\, x)\, V)=V\ ,\eqno(2.11)
$$
where $\approx$ denotes equality up to BRST-trivial terms.  This follows
because the second term on the right-hand side of (2.10), and the extra terms
that we incur by manipulating (2.10), with $A=V$, $B=x^{-1}$ and $C=x$, into
the  form (2.11), all involve commutators of physical operators, and these
commutators are BRST trivial by virtue of Lemma 2.1.  Using $(A\,
Q_B\xi)=Q_B(A\, \xi)$ for any physical operator $A$, the result then follows,
since $(x^{-1}\, (x\, V))$ can only yield the BRST non-trivial physical
operator $V$ if $(x\, V)$ is itself BRST non-trivial.  In other words $(x\,
V)$ must be non-trivial because we can act with the inverse of $x$ to get
back $V$ itself.  Clearly the proof works equally well for any invertible
physical operator acting on $V$, and thus in particular for $y$, $x^{-1}$
and $y^{-1}$ too.
\bigskip

     We have assumed in the above that the operators $x$, $y$, $x^{-1}$ and
$y^{-1}$ have well-defined normal-ordered products with any physical operator
in the theory.  To see that this is the case, we must show that the product
of the exponential operators, when normal ordered, gives an integer power of
$(z-w)$.  When $x$, with momentum $(8,\, 0)$, is normal ordered with a
physical operator with momentum given by $(k_1,\, k_2)$, this power will be
$-(\ft87\alpha)\, (\ft17 k_1\alpha)=-k_1$, which is clearly an integer.  When
$y$, with momentum given by $(4,\, 12)$, is normal ordered instead, the power
is $-(\ft47\alpha)\, (\ft17 k_1\alpha) -(\ft{12}7a)\, (\ft17 k_2
a)=-\ft12(k_1+k_2)$.  This is also always an integer, since any physical state
must have a momentum $(k_1,\, k_2)$ that satisfies (2.2) for an integer level
number $\ell$, and one can immediately see from (2.2) that $k_1$ and $k_2$ are
either both odd, or both even.

     With the above theorem, we are now able to construct the entire
cohomology of the BRST operator.   Since $x$ has momentum $(8,\, 0)$, and $y$
has momentum $(4,\, 12)$, it is clear that by acting with appropriate powers
of $x$ and $y$ on an arbitrary physical state $V'(k_1',k_2')$ with momentum
$(k_1',\, k_2')$, we can map it to a physical state $V(k_1,k_2)$ whose
momentum $(k_1,\, k_2)$ lies in a fundamental rectangular unit cell with
a width of 8 in the $k_1$ direction, and 12 in the $k_2$ direction.  For
reasons that will become clear presently, we choose our fundamental unit cell
to be the one defined by
$$
-8\le k_1\le -1,\qquad -11\le k_2\le 0\ .\eqno(2.12)
$$
By carrying out an exhaustive case-by-case analysis of the cohomology of all
states with momenta satisfying (2.12), we will therefore have solved the
complete cohomology problem for the two-scalar $W_3$ string, since all other
physical state in the theory can then be obtained by acting with all possible
integer powers of $x$ and $y$ on the fundamental physical states.

     Not all of the 96 lattice points in the fundamental cell (2.12) can
correspond to physical states, since $k_1$ and $k_2$ must satisfy the
mass-shell condition (2.2) for integer level numbers $\ell$.  By enumerating
the possibilities we find that there are just 12 solutions to (2.2) that
satisfy (2.12):
$$
\eqalign{
s_1&=(-6,\,-6)_0,\qquad s_2=(-6,\,-8)_0,\qquad s_3=(-8,\,-6)_0\ ,\cr
s_4&=(-8,\,-8)_0,\qquad s_5=(-7,\,-5)_0,\qquad s_6=(-7,\,-9)_0\ ,\cr
s_7&=(-8,\,0)_1,\qquad \ \ \; s_8=(-6,\,0)_1,\qquad\ \ \;s_9=(-4,\,-2)_1\ ,\cr
s_{10}&=(-3,\,-5)_1,\qquad s_{11}=(-3,\,-9)_1,\qquad s_{12}=(-2,\,-2)_2\ ,\cr}
\eqno(2.13)
$$
where the subscript on $(k_1,\, k_2)_\ell$ denotes the level number.  Since
the level numbers involved here are small, namely $\ell=0$, 1 and 2, it is
very easy to give a complete case-by-case analysis to find all of the
physical states specified by (2.13).  In fact the results are all already
known; the six $\ell=0$ states are tachyons, with ghost number $G=3$ [5,4]:
$$
\eqalign{
t_1&=c\,\del\gamma\,\gamma\, e^{-\ft67\alpha\varphi-\ft67 aX},\qquad
t_2=c\,\del\gamma\,\gamma\, e^{-\ft67\alpha\varphi-\ft87 aX},\qquad
t_3=c\,\del\gamma\,\gamma\, e^{-\ft87\alpha\varphi-\ft67 aX},\cr
t_4&=c\,\del\gamma\,\gamma\, e^{-\ft87\alpha\varphi-\ft87 aX},\qquad
t_5=c\,\del\gamma\,\gamma\, e^{-\alpha\varphi-\ft57 aX},\qquad\ \
t_6=c\,\del\gamma\,\gamma\, e^{-\alpha\varphi-\ft97aX},\qquad\,\cr}
\eqno(2.14)
$$
and the remaining six states, with $\ell=1$ and $\ell=2$,
have ghost number $G=2$ [7,8,4]:
$$
\eqalign{
u_1&=\big(c\,\gamma
+\ft{15}{14}\alpha\,\del\gamma\,\gamma\big) e^{-\ft87\alpha\varphi},\quad
\ \
u_2=\big(c\,\gamma -\ft37\alpha\,\del\gamma\,\gamma\big)
e^{-\ft67\alpha\varphi}\ ,\cr
u_3&=c\,\gamma\, e^{-\ft47\alpha\varphi-\ft27
aX},\qquad\qquad\ \ \
u_4=c\,\gamma\, e^{-\ft37\alpha\varphi-\ft57 aX}\ ,\cr
u_5&=c\,\gamma\, e^{-\ft37\alpha\varphi-\ft97 aX},\qquad\qquad\ \ \
u_6=c\big(\del\varphi\,\gamma-\ft37\alpha\, \del\gamma\big)
e^{-\ft27\alpha\varphi-\ft27 aX}\ . \cr}\eqno(2.15)
$$
These operators are the prime operators for each of the momenta listed in
(2.13). In other words, each of these is the lowest-ghost-number member of
a quartet of physical operators, of which the other three members are
$(a_\varphi\,V)$ and $(a_{\sss X}\,V)$ (boosted by ghost number
one),  and $(a_\varphi\,a_{\sss X} V)$ (boosted by ghost number two),
for  each prime operator $V$ [7,8,4]. The ghost boosters are defined by
$a_\varphi=[Q_B, \,\varphi]\ ,\ a_{\sss X}=[Q_B,\,X]$. Since the
processes of boosting and of normal ordering with $x^m\,y^n$ commute (up to
BRST trivial terms), we may always concentrate on the prime physical
operators, it being understood that each such operator is accompanied by its
three boosted partners.  Note that $x$, $y$, $x^{-1}$ and $y^{-1}$ themselves
are all prime physical operators, and that $u_1$ is the same operator as
$x^{-1}$.

      The discussion of the cohomology of the prime physical operators
divides into the odd-ghost-number case and the even-ghost-number case. For
odd ghost numbers, the prime physical operators are given by the
normal-ordered products
$$
x^m\,y^n\, t_i\ ,\eqno(2.16)
$$
with $G=3-2m-2n$, where $m$ and $n$ are arbitrary integers (positive or
negative) and $1\le i\le6$. Using (2.2) we find that their momenta and level
numbers are given by
$$
\eqalign{\phantom{V(m,n,1;G)} &\phantom{:}\quad \qquad\qquad
(k_1,k_2)\qquad\qquad \qquad\qquad\quad {\rm level\ number}\  \ell\cr
x^m\, y^n\, t_1&:\quad
(8m+4n-6,12n-6),\ \qquad 4m^2+4n^2+4m\,n+m+n\cr
x^m\, y^n\, t_2&:\quad (8m+4n-6,12n-8),\ \qquad
4m^2+4n^2+4m\,n+m\cr
x^m\, y^n\, t_3&:\quad (8m+4n-8,12n-6), \qquad
4m^2+4n^2+4m\,n-m\cr
x^m\, y^n\, t_4&:\quad (8m+4n-8,12n-8), \qquad
4m^2+4n^2+4m\,n-m-n\cr
x^m\, y^n\, t_5&:\quad (8m+4n-7,12n-5), \qquad
4m^2+4n^2+4m\,n+n\cr
x^m\, y^n\, t_6&:\quad (8m+4n-7,12n-9), \qquad
4m^2+4n^2+4m\,n-n \ .\cr}\eqno(2.17)
$$

For even ghost numbers, the prime operators are given by
$$
x^m\,y^n\, u_i\ ,\eqno(2.18)
$$
with $G=2-2m-2n$, where $m$ and $n$ are arbitrary integers and $1\le i \le 6$.
Again using (2.2) we find their momenta and level numbers  are given by
$$
\eqalign{\phantom{V(m,n,1;G)} &\phantom{:}\quad \qquad\qquad
(k_1,k_2)\qquad\qquad \qquad\qquad\quad {\rm level\ number}\  \ell\cr
x^m\,y^n\, u_1&:\quad
(8m+4n-8,12n),\phantom{-6}\ \qquad 4m^2+4n^2+4m\,n-m+3n+1\cr
x^m\,y^n\, u_2&:\quad (8m+4n-6,12n),\phantom{-6}\ \qquad
4m^2+4n^2+4m\,n+m+4n+1\cr
x^m\,y^n\, u_3&:\quad (8m+4n-4,12n-2), \qquad
4m^2+4n^2+4m\,n+3m+4n+1\cr
x^m\,y^n\, u_4&:\quad (8m+4n-3,12n-5), \qquad
4m^2+4n^2+4m\,n+4m+3n+1\cr
x^m\,y^n\, u_5&:\quad (8m+4n-3,12n-9), \qquad
4m^2+4n^2+4m\,n+4m+n+1\cr
x^m\,y^n\, u_6&:\quad (8m+4n-2,12n-2), \qquad
4m^2+4n^2+4m\,n+5m+5n+2 \ .\cr}\eqno(2.19)
$$
This completes the derivation of the entire cohomology of the two-scalar
$W_3$ BRST operator.

\np
%\bigskip
\noindent{\it 2.2 Further observations}
\bigskip

     It is interesting to note that the above construction automatically
gives the entire spectrum, which means that half of the quartets generated
by this procedure are conjugate to the other half. The momenta and ghost
numbers of the associated prime operators $V$ and $V'$ for a conjugate pair of
quartets are related by $k_1+k_1'=-14,\ k_2+k_2'=-14,\ G+G'=6$.  This last
equation follows because the operator conjugate to the prime operator $V$
is actually the twice-boosted operator $(a_\varphi\, a_{\sss X}\,
V')$ in the conjugate quartet, so that $(a_\varphi\, a_{\sss X}\,V')$
and $V$ together have the correct total ghost number 8.

     Another interesting observation is that although one might think that the
mass-shell condition (2.2) is only a {\it necessary} condition for the
occurrence  of physical operators, in fact there are physical operators
corresponding  to {\it every} pair of integers $(k_1,\,k_2)$ that satisfy (2.2)
for each integer $\ell$. To see this, note that if $(k_1,\,k_2)$ satisfies
(2.2) for some integer $\ell$, then $(k_1+8p,\,k_2+24q)$ also satisfies (2.2)
for some other integer $\ell'$, where $p$ and $q$ are arbitrary integers. Thus
if we can show that all solutions in one $8\times 24$ cell can be reached by
acting with powers of $x$ and $y$ on the twelve basic states (2.14) and
(2.15), then we have shown that all  solutions can be reached, since we then
can cover the whole $(k_1,\,k_2)$  plane by acting on the states in
the $8\times 24$ cell with $x^{p-q}\,y^{2q}$. There are  24 solutions to (2.2)
in this $8\times 24$ cell, and it is easy to show that they can indeed all be
obtained from the twelve fundamental states, thus completing the
demonstration.

     The mass-shell condition (2.2) is invariant under the action of the
Weyl group of $SU(3)$.  This can easily be seen by defining the shifted
momentum $(\hat k_1,\, \hat k_2)\equiv (k_1+7,\, k_2+7)$, in terms of which
(2.2) becomes
$$
3\hat k_1^2 +\hat k_2^2 =4(12\ell+1)\ .\eqno(2.20)
$$
By writing the simple roots of $SU(3)$ in the basis appropriate to our
conventions, one can show that the Weyl-group transformations corresponding
to the two simple roots are given by
$$
\eqalign{
S_1:&\qquad (\hat k_1,\, \hat k_2)\longrightarrow (\hat k_1,\, -\hat k_2)\ ,
\cr
S_2:&\qquad (\hat k_1,\, \hat k_2)\longrightarrow \big(\ft12(\hat k_2-\hat
k_1),\, \ft12(3\hat k_1+\hat k_2)\big)\ .\cr}\eqno(2.21)
$$
All six elements of the Weyl group can be generated from these.  Clearly,
these transformations leave (2.20) invariant.

     Since, as we have seen, there are physical states associated with all
solutions of (2.20), it follows that the action of the Weyl group on
(2.20) maps physical states at  any given level number into other physical
states at the same level.  For example, one easily sees that under (2.21),
the six $\ell=0$ tachyons (2.14) map into each other.  At higher levels,
however, the set of six Weyl-related physical states will be at different
ghost numbers.  An immediate consequence of this Weyl-group symmetry is that
the number $N$ of prime physical operators at each level $\ell$ is  necessarily
a multiple of 6.  It is not obvious how to determine $N$ as a function of
$\ell$ without simply enumerating the solutions to (2.2).  Some examples are
$\{\ell,\, N\}=\{0,\,6\},\  \{1,\,12\},\ \{2,\,6\},\  \{4,\,18\},\ \{7,\,0\},\
\{11,\,24\},\ \{53,\,36\},\ \{144,\,48\},\ \{690,\,54\}$.

     It will always be the case that exactly one member of each ``Weyl
sextet'' satisfies the ``Seiberg condition'' [9]
$$
\hat k_2 >0,\qquad \hat k_1> \ft13 \hat k_2\ ,\eqno(2.22)
$$
which singles out one of the six Weyl chambers.  As discussed in the
introduction, however, the condition (2.22) should not be interpreted as a
restriction on the validity of physical states in the free theory.  The
subset of the cohomology for which the physical states satisfy (2.22) was
constructed in [9].

    To conclude this section, we shall show that the calculation of the
normal-ordered product of any of $x$, $y$, $x^{-1}$ or $y^{-1}$ with a prime
physical operator $V$ can equivalently be performed by acting on a
ghost-boosted version of $V$ with the corresponding screening currents $S_x$,
$S_y$, {\it etc}.  From a practical point of view, this can often simplify the
calculation of higher-level operators considerably.  Thus we prove the
following

\medskip
\noindent{\bf Theorem 2.3}\hskip10pt {\it If $V$ is a prime physical
operator, then $[S_x\, (R\, V)]_1\approx (x\, V)$, and similarly for $y$,
$x^{-1}$ or $y^{-1}$, where $S_x$ is the screening current for $x$, defined
by (2.3), and $R$ is a certain linear combination of the $a_\varphi$ and
$a_{\sss X}$ ghost boosters.}

     Note that here, as in Lemma 2.1, $[A\, B]_r$ denotes the
coefficient of the $r$'th pole in the OPE of $A$ with $B$.  In particular,
$[A\, B]_0$ means the normal-ordered product of $A$ with $B$, which we
often also write as $(A\, B)$.  The expression
$[S_x\, (a\, V)]_1$ means $\oint dz\, S_x(z)\, (R\, V)(w)$, which is the
commutator of the screening charge $\oint dz\, S_x(z)$ with the boosted
physical operator.  As in Theorem 2.2, $\approx$  denotes equality up to
BRST-trivial terms.

     To prove the theorem, we use the following identity for any operators
$A$, $B$ and $C$ [14]:
$$
[A\, [B\, C]_0]_q=(-1)^{ab}[B\, [A\, C]_q]_0 + \sum_{r=0}^{q-1} {q-1 \choose
r} [[A\, B]_{q-r}\, C]_r\ ,\eqno(2.23)
$$
where $q\ge1$.  Thus we have
$$
[S_x\, (R\, V)]_1=[S_x\, [R\, V]_0]_1 =
-[R\, [S_x\, V]_1]_0 + [[S_x\, R]_1\, V]_0 \ .\eqno(2.24)
$$
Now from (2.3) we have $S_x=[b\,x]_1$, so $[S_x\, V]_1=[[b\, x]_1\, V]_1$ and
$[S_x\, R]_1=[[b\, x]_1\, R]_1$.  We may again use (2.23), with $q=2$, to
evaluate these expressions:
$$
\eqalign{
[[b\, x]_1\, V]_1=[b\, [x\, V]_0]_2-[x\, [b\, V]_2]_0 -[[b\, x]_2\, V]_0\ ,
\cr
[[b\, x]_1\, R]_1=[b\, [x\, R]_0]_2-[x\, [b\, R]_2]_0 -[[b\, x]_2\, R]_0\ .
\cr}\eqno(2.25)
$$
The first of these equations implies that $[[b\, x]_1\, V]_1\approx 0$.  This
is because $V$, $x$ and $(x\, V)$ are all {\it prime} physical operators, which
means that the BRST-closed operators $[b\, V]_2$, $[b\, x]_2$ and $[b\, [x\,
V]_0]_2$ ({\it i.e.}\ $b_0 V$, {\it etc.}\ )  must be BRST trivial, there
being, by the definition of a ``prime operator,'' no BRST-non-trivial physical
operator at the lower ghost number.  The second equation in (2.25) implies
that $[[b\, x]_1\, R]_1 \approx x$ (ignoring an unimportant non-zero constant
factor). This is because $R=({\rm const.})\,\del c + \cdots$, so $[b\,
R]_2={\rm const.}$, and $[b[x\,  R]_0]_2\approx ({\rm const.)} x$, since it is
annihilated by $Q_B$ and has  the same ghost number as the prime operator
$x$.  Substituting these results into (2.24), the theorem immediately follows.

     In order to see that $[[b\, x]_1\, R]_1$ above is a {\it non-vanishing}
constant multiple of $x$, which is essential to the proof, it is useful to
look more closely at the ghost boosters $a_\varphi\equiv [Q_B,\varphi]$ and
$a_{\sss X}\equiv [Q_B,X]$.  They are given by [4]
$$
\eqalign{
a_\varphi &=c\,\del\varphi -\alpha\,\del c -\ft{19}{8}\, \del^2\gamma
-\ft92\, \beta\,\del\gamma\,\gamma -3(\del\varphi)^2\,\gamma +3\alpha\,
\del\varphi\, \del\gamma\ ,\cr
a_{\sss X}&= c\, \del X -a\, \del c\ .\cr}\eqno(2.26)
$$
For calculational purposes, it is most convenient to choose the combination
$R$ of $a_\varphi$ and $a_{\sss X}$ in Theorem 2.3 so that $(R\, V)$
is annihilated by $b_0$, {\it i.e.}\ so that $[b\, (R\, V)]_2=0$.  It is
easy to see that the required combination is
$$
R=(k_2+7)a\, a_\varphi -(k_1+7)\alpha\, a_{\sss X}\ ,\eqno(2.27)
$$
where the momentum of the physical operator $V$ is given by $(k_1,\, k_2)$.
It is then straightforward to show that $[[b\, x]_1\, R]_1\approx \ft17 a\,
\alpha \big((k_1+7)\tilde k_2-(k_2+7)\tilde k_1\big)x$, where $(\tilde k_1,\,
\tilde k_2)$ is the momentum of $x$ (or, {\it mutatis mutandis}, $y$).
Elementary algebra using (2.2) then shows that the right-hand side of this
expression is always non-vanishing.

     Let us close with one sample calculation using the screening current
$S_x$ to build a higher-level physical operator.  We begin with the tachyon
$t_1$, which after boosting according to (2.27), becomes $c\, \del^2\gamma\,
\del\gamma\, \gamma\, e^{-\ft67\alpha\varphi-\ft67a X}$ (up to an irrelevant
constant factor).  It is convenient to bosonise the $(\beta,\gamma)$ ghosts,
so we write $\gamma=e^{i\rho}$ and $\beta=e^{-i\rho}$.  The boosted tachyon
then becomes $2\,c\, e^{3i\rho}\, e^{-\ft67\alpha\varphi -\ft67 a X}$.  The
screening current (2.4) bosonises to
$$
S_x= 2\,e^{-3i\rho}\, e^{\ft87\alpha\varphi}\ ,\eqno(2.28)
$$
and so $\oint dz\, S_x(z)$ acting on the boosted tachyon immediately gives
$$
4\oint {dz\over (z-w)^3}\, e^{-3i\rho(z)+\ft87\alpha\varphi(z)
+3i\rho(w)-\ft67\alpha\varphi(w)-\ft67 a X(w)}\eqno(2.29)
$$
after normal ordering the exponentials.  Thus re-expressing the result in
terms of $\beta$ and $\gamma$ again, we obtain the $\ell=5$ prime physical
operator
$$
\ft{16}7\alpha\, c\,\Big( 6\sqrt2\, \del\beta\, \gamma -3\sqrt2 \, \beta\,
\del\gamma +12\, \del\varphi\, \beta\,\gamma +4\sqrt2\, (\del\varphi)^2\,
+2\del^2\varphi\Big) e^{\ft27\alpha\varphi-\ft67 a X}\ .\eqno(2.30)
$$
This agrees with the result in [4], where it was found by directly solving
the physical-state conditions (1.9).  It should perhaps be emphasised,
however, that the constructions we are presenting in this paper are primarily
of utility for deriving general results about the spectrum of physical
states, rather than constructing specific explicit examples.  In practice, it
is easier to obtain specific examples by directlty solving the conditions
(1.6).

\bigskip
\noindent{\bf 3. Cohomology of the Multi-scalar $W_3$ String }
\bigskip

        The BRST operator for the multi-scalar realisation is related to
that for the two-scalar realisation in a very simple way, namely by
replacing the energy-momentum tensor (1.8) for $X$ by the one (1.7) for
$X^\mu$. Owing to this fact, the cohomology of the multi-scalar $W_3$ string
can be understood by examining the physical states of the two-scalar $W_3$
string.  In particular, the multi-scalar states that are tachyonic in the
effective spacetime described by $X^\mu$ can be obtained from the subset of
the states of the two-scalar $W_3$ string  that generalise to the
multi-scalar case. The remaining multi-scalar states can then be obtained by
replacing the tachyonic effective spacetime operator $e^{i p\cdot X}$ by
arbitrary excited highest-weight operators of the same conformal dimension
under $T^{\rm eff}$. Therefore the key problem is to identify the subset of
the states of the two-scalar $W_3$ string  that generalise to the
multi-scalar case.

         Physical operators in the two-scalar $W_3$ string will generalise
to multi-scalar operators if one of the following conditions holds:
\medskip

\item{1)} If there is a pair of two-scalar prime physical operators with
momenta $(k_1,\, k_2)$ and $(k_1,\, -14 -k_2)$, both at the same ghost
number. This pair, which have the same conformal dimension
$\Delta=-\ft1{48}(k_2+7)^2+\ft{49}{48}$ under $T^{\rm eff}$, generalise to a
continuous momentum multi-scalar operator where $e^{i p\cdot X}$ has the
same dimension $\Delta=\ft12 p^\mu(p_\mu+2a_\mu)$.

\item{2)} If there is a two-scalar prime physical operator with momentum
$(k_1,\, 0)$.  This generalises to a discrete multi-scalar operator with
$p_\mu=0$ in the effective spacetime, and hence $\Delta=0$.

\item{3)} If there is a two-scalar prime physical operator with momentum
$(k_1,\, -14)$.  This generalises to a discrete multi-scalar operator with
$p_\mu=-2a_\mu$ in the effective spacetime, where $a_\mu$ is the constant
background-charge vector appearing in (1.7).  Again, this has conformal
dimension $\Delta=0$ as measured by $T^{\rm eff}$.
\medskip

     By examining the entire cohomology of the two-scalar $W_3$ string,
given by (2.16)--(2.19), one can easily see that the subset of states that
fulfil these requirements can be described as follows:
$$
x^m\, t_i,\qquad x^m\, u_i,\qquad  x^m(x y^{-1} u_1),\qquad
x^m(x y^{-1} u_2),\qquad x^m(y^{-1}u_3), \qquad x^m(y^{-1} u_6)\ ,\eqno(3.1)
$$
where $m$ is an arbitrary integer in each case.  In fact
$$
\eqalign{
(x y^{-1} u_1)&=c\,\gamma\, e^{-\ft47\alpha\varphi-\ft{12}7aX}\ ,\cr
(x y^{-1} u_2)&=c\big(\del\varphi\,\gamma-\ft37\alpha\, \del\gamma\big)
e^{-\ft27\alpha\varphi-\ft{12}7 aX}\ .\cr}\eqno(3.2)
$$
In other words, these two operators are nothing but the $X^\mu$-space
momentum conjugates of $u_3$ and $u_6$ respectively.  The remaining
operators involving $y^{-1}$ in (3.2), namely $(y^{-1}u_3)$ and $(y^{-1}
u_6)$, have momenta $(-8,\,-14)$ and $(-6,\, -14)$; these are the $G=4$
prime physical operators for the quartets conjugate to those of $u_2$ and
$u_1$ respectively.

      There is an equivalent but more convenient way to describe the
cohomology (3.1) of the multi-scalar $W_3$ string, by starting with a set of
``basic'' physical operators that are expressed directly in the multi-scalar
formalism.  Thus we take the following basis of prime physical operators.
At $G=3$, there are [5,6,4]:
$$
\eqalign{
\tilde t_1&=c\,\del\gamma\,\gamma\, e^{-\ft67\alpha\varphi}\, V_1(X)\ ,\cr
\tilde t_2&=c\,\del\gamma\,\gamma\, e^{-\ft87\alpha\varphi}\, V_1(X)\ ,\cr
\tilde t_3&=c\,\del\gamma\,\gamma\, e^{-\alpha\varphi}\, V_{\ft{15}{16}}(X)\ ,
\cr}\eqno(3.3)
$$
where $V_\Delta(X)$ denotes an effective-spacetime physical operator, which
is highest-weight under $T^{\rm eff}$ with conformal dimension $\Delta$.
At $G=2$, there are further prime operators with continuous spacetime
momentum [7,8,4]:
$$
\eqalign{
\tilde u_1&=c\, \gamma\, e^{-\ft37\alpha\varphi}\, V_{\ft{15}{16}}(X)\ ,\cr
\tilde u_2&=c\,\gamma\, e^{-\ft47\alpha\varphi}\, V_{\ft12}(X)\ ,\cr
\tilde u_3&=c\big(\del\varphi\,\gamma
-\ft37\alpha\,\del\gamma\big)e^{-\ft27\alpha\varphi}\,V_{\ft12}(X)\ .\cr}
\eqno(3.4)
$$
Finally, there are discrete prime operators $\tilde d_1$ and $\tilde d_2$ at
$G=2$ [7], and $\tilde d_3$ and $\tilde d_4$ in the conjugate quartets, at
$G=4$:
$$
\eqalign{
\tilde d_1&= \big(c\, \gamma +\ft{15}{14}\alpha\,\del\gamma\,\gamma\big)
e^{-\ft87\alpha\varphi}\ ,\cr
\tilde d_2&= \big(c\, \gamma -\ft37\alpha\,\del\gamma\,\gamma\big)
e^{-\ft67\alpha\varphi}\ ,\cr
\tilde d_3&= c \big(\del\varphi\,\del^2\gamma\,\del\gamma\, \gamma
-\ft{4}{21}\alpha\,\del^3\gamma\,\del\gamma\,\gamma\big)
e^{-\ft67\alpha\varphi-2i a\cdot X}\ , \cr
\tilde d_4&= c \big(\del\varphi\,\del^2\gamma\,\del\gamma\, \gamma
-\ft17\alpha\,\del^3\gamma\,\del\gamma\,\gamma\big)
e^{-\ft87\alpha\varphi-2i a\cdot X}\ ,\cr}\eqno(3.5)
$$

     From the basic prime operators (3.3)--(3.5), all the prime physical
operators of the multi-scalar $W_3$ string are obtained, by normal ordering
them with $x^m$ for arbitrary integers $m$.  (Note that both $x$ and
$x^{-1}$ themselves generalise to the multi-scalar case, since they have zero
momentum in the $X^\mu$ directions.)  Thus the complete spectrum of
prime physical operators in the multi-scalar $W_3$ string is as follows:

\np

%\bigskip
\settabs 5 \columns
\+Operator & $k_1$ & $G$ & $\Delta$ & $\ell$ \cr
\medskip

\+$x^m\, \tilde t_1$ & $8m-6$ & $3-2m$ & $1$ & $4m^2+m$ \cr
\medskip

\+$x^m\, \tilde t_2$ & $8m-8$ & $3-2m$ & $1$ & $4m^2-m$ \cr
\medskip

\+$x^m\, \tilde t_3$ & $8m-7$ & $3-2m$ & $\ft{15}{16}$ & $4m^2$ \cr
\medskip

\+$x^m\, \tilde u_1$ & $8m-3$ & $2-2m$ & $\ft{15}{16}$ & $4m^2+4m+1$ \cr
\medskip

\+$x^m\, \tilde u_2$ & $8m-4$ & $2-2m$ & $\ft12$ & $4m^2+3m+1$ \cr
\medskip

\+$x^m\, \tilde u_3$ & $8m-2$ & $2-2m$ & $\ft12$ & $4m^2+5m+2$ \cr
\medskip

\+$x^m\, \tilde d_1$ & $8m-8$ & $2-2m$ & $0$ & $4m^2-m+1$ \cr
\medskip

\+$x^m\, \tilde d_2$ & $8m-6$ & $2-2m$ & $0$ & $4m^2+m+1$ \cr
\medskip

\+$x^m\, \tilde d_3$ & $8m-6$ & $4-2m$ & $0$ & $4m^2+m+1$ \cr
\medskip

\+$x^m\, \tilde d_4$ & $8m-8$ & $4-2m$ & $0$ & $4m^2-m+1$ \cr
\medskip
\centerline{\it Table 1.\ \ \ Prime physical operators in the multi-scalar
$W_3$ string}
\bigskip

The level numbers $\ell$ in the table are for the case where the
effective-spacetime highest-weight operators $V_\Delta(X)$ are purely
tachyonic, $V_\Delta(X)=e^{i p\cdot X}$.  The total level number of any
physical operator would be given by the sum of $\ell$ and the level number
of the excitation number of $V_\Delta(X)$.  From Table 1, we see that
physical operators in the $\Delta=1$ sector occur at levels $\ell=0,\, 3,\,
5,\,14,\,18,\,33,\,39,\ldots$; in the $\Delta=\ft{15}{16}$ sector at
$\ell=0,\,1,\,4,\,9,\,16,\,25,\,36\,\ldots$; and in the $\Delta=\ft12$ sector
at $\ell=1,\,2,\,8,\,11,\,23,\,28,\,46,\ldots$.  The discrete physical
operators in the $\Delta=0$ sector occur at levels $\ell=1,\,4,\,6,\,15,\,
19,\,34,\,40,\ldots$.

     Note that for all six sequences of physical operators $x^m\, \tilde t_i$
and $x^m\, \tilde u_i$, the set with $m<0$ correspond to the conjugates of the
set with $m\ge0$.  These all have continuous on-shell spacetime momentum
$p_\mu$, subject only to the mass-shell condition $\Delta=\ft12 p^\mu(p_\mu+2
a_\mu)$, with intercepts $\Delta$ as given in the first six lines of Table 1
above.  All of the discrete operators $x^m\,\tilde d_1$ and $x^m\, \tilde d_2$
have $p_\mu=0$, and all of the discrete operators  $x^m\,\tilde d_3$ and
$x^m\, \tilde d_4$ have $p_\mu=-2\, a_\mu$. The complete cohomology of
physical operators that we find here in the multi-scalar $W_3$ string accords
with the pattern that was observed in many low-level examples in [8,4].
However, the discrete operators $x^m\,\tilde d_3$  and $x^m\, \tilde d_4$ with
spacetime momentum $p_\mu=-2\, a_\mu$ were not found previously.

      One can easily verify that the subset of the two-scalar cohomology
(2.16)--(2.19) that satisfies any of the three criteria for generalisability
given at the beginning of this section is the same as the multi-scalar
cohomology given in the table above, where one restricts the
effective-spacetime highest-weight operators $V_\Delta(X)$ to be purely
tachyonic.  Many, indeed the majority, of the two-scalar operators are
``lost'' when one passes to the multi-scalar $W_3$ string.  The reason for
this is that the $y$ operator of the two-scalar $W_3$ string does not
generalise to the multi-scalar case, and so it can no longer play a r\^ole  in
generating higher-level physical operators from the basic $G=3$ and $G=2$
physical operators.   Thus whilst the entire cohomology of the two-scalar
$W_3$ string is generated by acting with all possible integer powers of the
$G=-2$ operators $x$ and $y$ on a basis of $G=3$ and $G=2$ operators, in the
multi-scalar $W_3$ string the entire cohomology is generated by acting with
all possible powers of just the $x$ operator on a basis of $G=3$ and $G=2$
continuous-momentum operators, and $G=4$ and $G=2$ discrete operators.  A
consequence of this is that at any given level number $\ell$, the range of
ghost numbers at which prime physical operators occur is much sparser in the
multi-scalar case than in the two-scalar case.  In fact in the multi-scalar
$W_3$ string, if continuous momentum prime operators occur at all at a given
level, then they occur at just two ghost numbers; $G=g$ and $6-g$
(corresponding to a quartet and its conjugate), where $g$ becomes increasingly
negative as $\ell$ increases.  If discrete prime operators occur at a given
level, then they occur at four ghost numbers; $G=g$, $g+2$, $4-g$ and $6-g$
(corresponding to two quartets\footnote{$^*$}{\tenfoot Since in the
multi-scalar $W_3$ string there are $(d+1)$ ghost boosters  $a_\varphi$ and
$a_{{\sss X}^\mu}$, the structure of boosted  operators can be more
complicated than simply quartets.  As discussed in  [4], one still gets
quartets from prime operators of the form $c\,  U(\varphi,\beta,\gamma)\,
e^{ip\cdot X}$, but a discrete prime operator at  $G=g$ with $p_\mu=0$, which
has the form $V=c\, U_1(\varphi,\beta,\gamma) +  U_2(\varphi,\beta,\gamma)$,
gives rise to a multiplet at ghost numbers $\{  g,\, g+1,\, g+2,\, g+3\}$,
with multiplicities $\{1,\, d+1,\, 2d-1,\,  d-1\}$.  It is convenient to view
these as a standard quartet, generated by  $a_\varphi$ and
$a^{\scriptscriptstyle /\!/}$ acting on the prime operator $V$, and a  set of
$(d-1)$ further quartets, spanning the ghost numbers $\{g+1,\, g+2,\,  g+3\}$,
generated by acting with these same two boosters on the $(d-1)$ operators
$a^\perp_{{\sss X}^\mu}V$ at $G=g+1$.  ($a^{\scriptscriptstyle  /\!/}$
denotes $a_{{\sss X}^\mu}$ projected parallel to the background-charge
vector $a_\mu$, and $a^\perp_{{\sss X}^\mu}$ denotes the remaining
$a_{{\sss X}^\mu}$ ghost boosters perpendicular to $a_\mu$.)  These
$(d-1)$ quartets  can alternatively be viewed as special cases of the
$\Delta=1$ continuous  momentum physical operators that always occur at the
next-lower level number, in which the spacetime tachyon $e^{ip\cdot X}$ is
replaced by excited $\Delta=1$ operators of the form $\xi_\mu\,\del
X^\mu$.} and their conjugates).  In the two-scalar $W_3$ string, by contrast,
prime operators can occur at many of the ghost numbers $G=g$, $g+2$,
$g+4$,$\cdots$ $2-g$, $4-g$, $6-g$.  (It can sometimes also happen that even
and odd ghost-number sequences occur at the same level.)  For example, at level
$\ell=14$, the two-scalar $W_3$ string has 18 prime operators, at ghost
numbers $G=-1$, 1, 3, 5 and 7; in the multi-scalar case, there are just two
prime operators, with $\Delta=1$ and $G=-1$ and $G=7$.  At level $\ell=46$,
there are 24 prime operators in the two-scalar $W_3$ string, at ghost numbers
$G=-5$, $-4$, $-1$, 2, 4, 7, 10 and 11; in the multi-scalar case there are two
prime operators, with $\Delta=\ft12$ and $G=-4$ and 10.  At level $\ell=30$,
there are 18 prime operators in the two-scalar $W_3$ string, at ghost numbers
$G=-4$, $-3$, $-1$, 0, 1, 5, 6, 7, 9 and 10; in the multi-scalar case, there
are no physical operators at this level.

\bigskip
\noindent{\bf 4. The Cohomology of the One-scalar String}
\bigskip

     The methods that we have used to study the cohomology of the $W_3$
string can also be applied to the simpler problem of the one-scalar Virasoro
string.  The BRST operator in this case is simply given by
$$
Q_B=\oint dz\, c\,\big( T + \ft12 T_{c,b}\big)\ ,\eqno(4.1)
$$
where
$$
\eqalignno{
T&=-\ft12 (\del\varphi)^2 -\alpha\, \del^2\varphi\ ,&(4.2)\cr
T_{c,b}&= -2\, b\, \del c - \del b\, c\ , &(4.3)\cr}
$$
where the background charge parameter $\alpha$ now satisfies $\alpha^2=
\ft{25}{12}$.  Physical operators have momenta $p=\ft15 k\, \alpha$,
where $k$ is an integer, which implies that the mass-shell condition takes the
form $$
(k+5)^2=24\ell+1\ ,\eqno(4.4)
$$
where $\ell$ is the level number.  Thus $k$ is always an even integer, and
physical operators can occur at levels $\ell=0,\, 1,\, 2,\, 5,\, 7,\, 12,\,
15,\, 22,\ldots$.

     We find that there is a prime physical operator at $\ell=5$ with  momentum
given by $k=6$.  This means that it has a well-defined normal-ordered product
with all physical operators.  It has ghost number $G=-1$, and takes the form
$$
x=\Big(-\ft12 \del\varphi\, \del b-\sqrt3\, \del b\, b\, c +\del^2\varphi\, b
+\ft1{2\sqrt{3}}\, \del^2 b\Big)e^{\ft65\alpha\varphi}\ .\eqno(4.5)
$$
The inverse of this operator occurs at $\ell=0$ with $k=-6$ and $G=1$:
$$
x^{-1} =c\,e^{-\ft65\alpha\varphi}\ .\eqno(4.6)
$$
Similar arguments to those in section 2 imply that $x^m$ acting on any
physical operator gives another BRST-non-trivial physical operator. Thus the
cohomology of the one-scalar string may be determined by finding all
physical operators whose momenta lie within an interval $n\le k\le n+5$ for
any convenient integer $n$. We shall choose $n=-6$.  From (4.4), it follows
that the only physical operators with $-6\le k\le -1$ are the $\ell=0$
tachyons, which have $G=1$:
$$
t_1=c\, e^{-\ft45\alpha \varphi},\qquad t_2=c\, e^{-\ft65\alpha\varphi}\ .
\eqno(4.7)
$$
Thus the complete cohomology of prime physical operators in the one-scalar
string is given by
$$
x^m\, t_1,\qquad x^m\, t_2\ ,\eqno(4.8)
$$
where $m$ is an arbitrary integer.  The momenta, ghost numbers and level
numbers are given by:

\bigskip
\+ & Operator & $k$ & $G$ & $\ell$ \cr
\medskip

\+ & $x^m\, t_1$ & $6m-4$ & $1-m$ & $\ft12 m(3m+1)$\cr
\medskip

\+ & $x^m\, t_2$ & $6m-6$ & $1-m$ & $\ft12 m(3m-1)$\cr
\bigskip
\centerline{\it Table 2. \ \ Prime physical operators in the one-scalar
string}
\bigskip

Each of the prime operators given above is associated with a doublet of
physical operators, where the second member is obtained from the prime
operator by normal ordering with the ghost booster $a_\varphi= c\,\del
\varphi -\alpha\, \del c$.  This gives the complete cohomology of the
one-scalar string.  One can verify that it agrees with the results in [15,16]
for  the two-scalar Virasoro string when the central charge of the matter field
$X$ is chosen to be zero.

%\np
\bigskip
\noindent{\bf 5. Discussion and Conclusions }
\bigskip

        In this paper, we have studied the complete spectrum of physical
states in the two-scalar and multi-scalar $W_3$ strings and in the
one-scalar Virasoro string. In all of these theories, there exist certain
special physical operators that are invertible, and by normal ordering
arbitrary powers of these operators with a set of basic operators, all
physical operators can be constructed.

     In earlier work on the spectrum of the $W_3$ string, it was found
that there are two $G=0$ prime operators at level $\ell=6$ in the two-scalar
$W_3$ string, which we shall call $\tilde x$ and $\tilde y$, with momenta given
by $(k_1,\, k_2)=(2,0)$ and $(1,3)$ respectively [7].  The operator $\tilde x$
generalises to the multi-scalar $W_3$ string, whilst $\tilde y$ does not.  In
terms of the construction in this paper, they are given by $\tilde x=x\, u_2$
and $\tilde y=y\, u_5$ respectively.  It was suggested in [7] that one might
use these operators in order to build higher-level physical states from a
basis of low-level states. In fact, because the trend is for the ghost numbers
of prime operators to cover a wider and wider range of values as the level
number increases, one would need also to use the $G=-1$ screening currents
built from $\tilde x$ and $\tilde y$ by inserting $\oint dz\, b(z)$
operators.  Some examples were presented in [17] for the special case of the
multi-scalar $W_3$ string.  There are many disadvantages to trying to use the
$\tilde x$ and $\tilde y$ operators for building up the entire cohomology of
the theory.  First of all, these operators do not have inverses, which means
that it is difficult to be sure that the higher-level states that one builds
are BRST non-trivial.  Indeed in general, unless one adjusts the number of
insertions of $\oint dz\, b(z)$ carefully, the result will certainly be BRST
trivial, since it will have a ghost number at which no non-trivial states
exist.  Only if the ghost numbers and momenta of the physical operators are
already known from some other argument does one know how to make the
appropriate number of $\oint dz\, b(z)$ insertions.  Even then, it is not
obvious that the physical state that one arrives at will necessarily be BRST
non-trivial.  Another related difficulty with using the $\tilde x$ and $\tilde
y$ operators is that they do not in general give integer-degree poles when
normal ordered with other physical states, or, indeed, with themselves.  This
leads to  tedious complications in performing the multiple contour integrals
that arise when building higher-level states.  It was observed in [7] that the
fourth powers of $\tilde x$ and $\tilde y$ {\it do} always give integer
poles.  The reason for this is that their momenta correspond to those of the
$x$ and $y$ operators at $\ell=15$.  Indeed, all of the above-mentioned
difficulties are avoided by using the $\ell=15$ operators $x$ and $y$, as we
have seen in this paper.

     One of the outstanding problems for $W_3$ strings is to understand the
``$W_3$ geometry'' on the worldsheet, and its r\^ole in governing the
structure of the physical spectrum and the interactions.  In some sense the
multi-scalar $W_3$ string is somewhat trivial, in that the physical
states all factorise into products of effective-spacetime states with primary
fields of the Ising model realised by the $(\varphi,\beta,\gamma)$ system
[5,6,8,4].  The r\^ole of the $W_3$ symmetry is rather trivial in this case.
However, it may be that the much richer spectrum of the two-scalar $W_3$
string is associated with a more non-trivial action of the symmetry.
Possibly this can be related to the discussion of $W_3$ geometry given in
[18].

%\np
\bigskip
\bigskip
\singlespace
\centerline{\bf REFERENCES}
\frenchspacing
\bigskip

\item{[1]}J. Thierry-Mieg, {\sl Phys. Lett.} {\bf B197} (1987) 368.

\item{[2]}V.A. Fateev and A.B. Zamolodchikov, {\sl Nucl. Phys.} {\bf B280}
(1987) 644.

\item{[3]}L.J.  Romans, {\sl Nucl.  Phys.} {\bf B352} (1991) 829.

\item{[4]}H. Lu, C.N. Pope, S. Schrans and X.J. Wang,  ``On the spectrum
and scattering of $W_3$ strings,'' preprint CTP TAMU-4/93, KUL-TF-93/2,
hep-th/9301099, to appear in Nucl. Phys. B.

\item{[5]}S.R. Das, A. Dhar and S.K. Rama, {\sl Mod. Phys. Lett.}
{\bf A6} (1991) 3055; {\sl Int. J. Mod. Phys.} {\bf A7} (1992) 2295.

\item{[6]}C.N. Pope, L.J. Romans, E. Sezgin and K.S. Stelle,
{\sl Phys. Lett.} {\bf B274} (1992) 298.

\item{[7]}C.N. Pope, E. Sezgin, K.S. Stelle and X.J. Wang, {\sl Phys. Lett.}
{\bf B299} (1993) 247.

\item{[8]}H. Lu, C.N. Pope, S. Schrans and X.J. Wang, {\sl Nucl. Phys.} {\bf
B403} (1993) 351.

\item{[9]}P. Bouwknegt, J. McCarthy and K. Pilch, ``Semi-infinite
cohomology of $W$ algebras,'' USC-93/11, hep-th/9302086.

\item{[10]}A. Andersen, B.E.W. Nilsson, C.N. Pope and K.S. Stelle, to appear.

\item{[11]}E. Witten, {\sl Nucl. Phys.} {\bf B373} (1992) 187;\nl
E. Witten and B. Zwiebach, {\sl Nucl. Phys.} {\bf B377} (1992) 55.

\item{[12]}See, for example, [6], and H. Lu, C.N. Pope, S. Schrans and K.W.
Xu,  {\sl Nucl. Phys.} {\bf B385} (1992) 99.

\item{[13]}K. Thielemans, {\sl Int. J. Mod. Phys.} {\bf C2} (1991) 787.

\item{[14]}F. Bais, P. Bouwknegt, M. Surridge and K. Schoutens, {\sl Nucl.
Phys} {\bf B304} (1988) 348.

\item{[15]}B.H. Lian and G.J. Zuckerman, {\sl Phys. Lett.} {\bf 254B} (1991)
417; {\sl Phys. Lett.} {\bf 266B} (1991) 21; {\sl Comm. Math. Phys.} {\bf
145} (1992) 561.

\item{[16]}P. Bouwknegt, J. McCarthy and K. Pilch, {\sl Comm. Math. Phys.}
{\bf 145} (1992) 541.

\item{[17]}M.D. Freeman and P.C. West, ``The covariant scattering and
cohomology of $W_3$ strings,'' preprint, KCL-TH-93-2, hep-th/9302114.

\item{[18]}J.-L. Gervais and Y. Matsuo, {\sl Comm. Math. Phys.} {\bf 152}
(1993) 317.

\end